\documentclass[12pt,preprint]{aastex}
\usepackage{amsmath}
\usepackage{ccaption}
\usepackage{color}
\usepackage{graphicx}
\usepackage{subfigure}
\newcommand{\dosuu}{{}$^\circ$}
\newcommand{\HeI}{He\,\textsc{i}}
\newcommand{\NII}{N\,\textsc{ii}}
\newcommand{\OI} {O\,\textsc{i}}
\newcommand{\OII}{O\,\textsc{ii}}
\newcommand{\OIII}{O\,\textsc{iii}}
\newcommand{\SII}{S\,\textsc{ii}}
\newcommand{\CaII}{Ca\,\textsc{ii}}
\newcommand{\MgII}{Mg\,\textsc{ii}}
\newcommand{\lam}[1]{\ensuremath{\lambda}\,#1}
\newcommand{\lamlam}[1]{\ensuremath{\lambda\lambda}\,#1}
\newcommand{\csalt}{$c_{salt}$}
\newcommand{\dm}{$\Delta\!m_{15}$}
\newcommand{\ha}{\rm{H}$\alpha$}
\newcommand{\hb}{\rm{H}$\beta$}
\newcommand{\sib}{\rm{Si}\,\textsc{ii}~{\lam 5972}}
\newcommand{\sic}{\rm{Si}\,\textsc{ii}~{\lam 6355}}

\newcommand{\xx}{$x_1$}
\newcommand{\tmax}{$t_{max}$}
\newcommand{\vel}[1]{#1 km~sec$^{-1}$}
\newcommand{\ebv}{$E(B-V)$}
\newcommand{\IRAF}{{\rm IRAF}}

\shorttitle{SDSS-II Subaru Spectra}
\shortauthors{Konishi et al.}

\begin{document}

\title{Subaru Spectroscopy of SDSS-II Supernovae \altaffilmark{1}}

\author{Kohki Konishi\altaffilmark{2,3}, 
Naoki Yasuda\altaffilmark{3,4},
Kouichi Tokita\altaffilmark{5},
Mamoru Doi\altaffilmark{5},
Yutaka Ihara\altaffilmark{5},
Tomoki Morokuma\altaffilmark{6},
Naohiro Takanashi\altaffilmark{7},
Jakob Nordin\altaffilmark{8,9},
John Marriner \altaffilmark{10},
Linda \"{O}stman\altaffilmark{11},
Michael Richmond\altaffilmark{12},
Masao Sako\altaffilmark{13},
Donald P. Schneider\altaffilmark{14},
J. Craig Wheeler\altaffilmark{15}
}
\email{kohki@icrr.u-tokyo.ac.jp}

\altaffiltext{1}{Based in part on data collected at Subaru Telescope, which is operated by the National Astronomical Observatory of Japan.}
\altaffiltext{2}{Department of Physics, Graduate School of Science, University of Tokyo, Tokyo 113-0033, Japan}
\altaffiltext{3}{Institute for Cosmic Ray Research, University of Tokyo, Kashiwa 277-8582, Japan}
\altaffiltext{4}{Institute for the Physics and Mathematics of the Universe, University of Tokyo, Kashiwa 277-8582, Japan}
\altaffiltext{5}{Institute of Astronomy, Graduate School of Science, University of Tokyo, 2-21-1 Osawa, Mitaka, Tokyo 181-0015, Japan}
\altaffiltext{6}{National Astronomical Observatory of Japan, 2-21-1, Osawa, Mitaka, Tokyo, 181-8588, Japan}
\altaffiltext{7}{Institute of Industorial Science, University of Tokyo, 153-8505, Japan}
\altaffiltext{8}{Department of Physics, Stockholm University, 106 91 Stockholm, Sweden.}
\altaffiltext{9}{Oskar Klein Centre for Cosmo Particle Physics, AlbaNova, 106 91 Stockholm, Sweden.}
\altaffiltext{10}{Center for Particle Astrophysics, Fermi National Accelerator Laboratory, Batavia, IL 60510, USA}
\altaffiltext{11}{Institut de F\'{i}sica d'Altes Energies, Universitat Aut\`{o}noma de Barcelona, Bellaterra, Spain}
\altaffiltext{12}{Physics Department, Rochester Institute of Technology, Rochester, NY 14623, USA}
\altaffiltext{13}{Department of Physics and Astronomy, University of Pennsylvania, 209 South 33rd Street, Philadelphia, PA 19104, USA}
\altaffiltext{14}{Department of Astronomy and Astrophysics, Pennsylvania State University, University Park, PA 16802, USA}
\altaffiltext{15}{Astronomy Department, University of Texas, Austin, TX 78712, USA}

\begin{abstract}
The Sloan Digital Sky Survey II (SDSS-II) Supernova Survey discovered Type Ia supernovae (SNe Ia) in an almost unexplored intermediate redshift range of $0.05 < z < 0.4$ and provided densely sampled multi-color light curves for SN candidates. 
Followup spectroscopy of this survey was carried out with the Subaru telescope and spectra of 71 SN Ia candidates were obtained.  One spectrum was observed per candidate except for a peculiar variable. This paper presents the method for processing these spectra. 
The observed wavelength ranges of our spectra are 4000 to 9000 {\AA} for Year 2005 and 3600 to 9000 {\AA} for Year 2006.  Most SN Ia spectra have signal to noise ratios (S/N) between 4 and 10 per 2 {\AA} averaged over the entire wavelength region. 
We developed a new code to extract the SN spectral component from spectra contaminated by the host galaxy.
Of 71 SN Ia candidates, 59 are identified as normal SNe Ia and 3 are peculiar SNe Ia. The range of spectral phases varies from -7 days to +30 days from maximum brightness.
There are also 7 SNe II, 1 possible hypernova and 1 AGN.
\end{abstract}


\section{Introduction}
Type Ia Supernovae (SNe Ia) have played a significant role in observational
cosmology as a probe for the expansion history of the Universe \citep[e.g.]
[]{rie98,per99}.
This is because the majority of the observed SNe Ia have homogeneous 
light curves and spectra. 

Recent nearby observations have enabled the construction of large datasets of SNe Ia obtained by a homogeneous
procedure.
Many spectra have been stored in the SUpernova SPECTra (SUSPECT) database \footnote{\url{http://bruford.nhn.ou.edu/\~{}suspect/}}.
A supernova group in the Center for Astrophysics (CfA) presented UBVRI photometry of 44 nearby SNe Ia \citep{jha06} and 185 SNe Ia \citep{hic09a} and 32 optical spectra \citep{mat08}.
The Carnegie Supernova Project recently published $ugri$BVYJHK$_s$ light curves of 35 nearby SNe Ia \citep{con09}. 
The Nearby Supernova Factory \citep{ald02} has been gathering spectrophotometry data of nearby SNe Ia.
High redshift (high-z) SNe Ia have been observed by several groups: 
the High-Z team \citep{rie98}, the Supernova Cosmology Project \citep{per99}, the SuperNova Legacy Survey \citep{ast06} and the Equation of State: SupErNovae trace Cosmic Expansion survey \citep{mik07}. However, since wide field observations are required to obtain comparable numbers of SNe Ia, few have been observed in the intermediate redshift range ($0.05<z<0.4$).

The SDSS-II Supernova Survey carried out a three year project \citep{fri08} to obtain multi-band light curves of SNe Ia at intermediate redshifts.
The SDSS 2.5m telescope \citep{gun06} scanned 300 square deg with the SDSS imaging camera \citep{gun98} and five filters \citep[$ugriz$, ][]{fuk96}.
The observed cadence was two days minimum for the same sky region.
Analysis was conducted in real-time to assign SN candidates for spectroscopic observations \citep{sak08}.
Follow-up spectroscopic observations were carried out at a large number of telescopes to determine at least the supernova type and redshift; the 2.4m Hiltner, the 2.5m NOT, the 3.5m SDSS ARC, the 3.6m NTT, the 4.2m WHT, the 8.2m Subaru, the 9.2m HET and the 10m Keck I telescopes. The spectroscopy of the first season (Year 2005) is presented in \citet{zhe08}. \citet{ost10} present the spectra obtained at the NTT and the NOT during Year 2006 and 2007. \citet{hol08} provide high-quality SN light curves in $ugriz$ for the first season.
\citet{fol10} describes the Keck followup spectra and presents a mismatch in the ultraviolet spectra of these SNe Ia from nearby counterparts.
The final data release for all spectra obtained by these telescopes will be 
published in Sako et al (in preparation).

In this paper, we present an observational summary of the Subaru spectroscopic followup. 
\S \ref{observ} describes our observations.
\S \ref{reduc} describes spectroscopic data reduction including the
method to extract SN spectra from host galaxy contaminated spectra.
\S \ref{datasets} describes the dataset obtained.
We summarize this paper in \S \ref{summary}.

\section{Observations}\label{observ}

We used the Subaru telescope for follow-up spectroscopic observations 
of the SN candidates found by the SDSS-II Supernova Survey for the 
first two seasons (Years 2005 \& 2006). 
The Faint Object Camera and Spectrograph \citep[FOCAS;][]{kas02} was 
used in the long-slit mode.
The atmospheric dispersion corrector (ADC) was used in our observations.
The instrumental dispersions were 1.34 {\AA}/pix.
We used a 0.8'' slit for SN candidates and a 2.0'' slit for 
spectroscopic standard stars.
The pixel scale was 0.104'' and the data were read out with three 
pixel binning in the spatial direction.
In order to obtain spectra in as wide a wavelength range as possible, we observed red and blue sides separately with different sets of grisms and filters. 
We obtained a red side spectrum first to identify the ``{\sic}'' feature which characterize SNe Ia and then a blue side spectrum unless 
the weather was bad or the observing time was limited. 
SN candidates were observed at the time when they 
were as close to zenith as possible and above an elevation of 
30{\dosuu}, so that they suffered the least from atmospheric absorption.

For Year 2005, we adopted the following grating and filter combinations  for the blue and the red sides; 
(i) 300 line mm$^{-1}$ grating blazed at 5500 {\AA} (300B) and the L600 filter, which is sensitive in the wavelength window of 4000 to 6000 {\AA}, and 
(ii) 300 line mm$^{-1}$ grating blazed at 7500 {\AA} (300R) and the SY47 filter, which is sensitive in the wavelength window of 5000 to 9000 {\AA}. 
The configuration for Year 2006 was the same as the first year except the filter for the blue side.
We replaced the L600 filter with a L550 filter which is more efficient for shorter wavelengths, widening the spectral window to 3600 to 9000 {\AA}.
Table \ref{setups} summarizes our configurations.

Because of the Subaru's large aperture and a superb seeing site, we observed the subset of SDSS targets with large host galaxy contamination. 
We aligned the slit so as to go through each SN candidate and host center.  
The slit angles were initially determined from the images taken by the SDSS 
2.5m telescope, but then adjusted by checking the $r$-band images taken by 
the Subaru telescope at the time of each observation.  This slit alignment 
enabled us to simultaneously determine the SN type as well as its redshift.  
The redshift was generally determined using spectral lines of the host galaxy 
(see \S \ref{zz} for details).  The observations of the SN candidates were 
not typically made at the parallactic angle, and the observed spectra should 
be color corrected due to differential atmospheric refraction when the 
airmass was large. The atmospheric refraction was corrected by the ADC. 89 \% 
of our targets were observed at an airmass smaller than 1.5. 

We usually observed two spectrophotometric standard stars each night for the
flux calibration; at the beginning and the end of observations.
Table \ref{stdstar} summarizes the list of standard stars observed 
in Years 2005 and 2006.
All the SN candidates were observed for $3 \times 300$ seconds for both
the blue and red sides.  Candidates spectra were observed only once.  
The only exception was a peculiar variable SN10450.
We observed SN10450 with an exposure of 900 seconds and 1,800 seconds 
(\S \ref{zz}, \citet{tok09}). 

We determined types and redshifts of each target via real-time
analysis.
This information was collected by our spectroscopic coordinator and
reported to the International Astronomical Union (IAU).
Type identification of three targets in our list was delayed due to
high contamination by the host galaxy and lower S/N.
Detailed off-line analysis identified these three targets as SNe Ia.

\section{Spectra reduction} \label{reduc} 

The spectroscopic data were reduced using the Image Reduction and
Analysis Facility ({\IRAF}) 
\footnote{{\IRAF} is distributed by the National Optical 
Astronomy Observatories, which are operated by the Association of
Universities for Research in Astronomy, Inc., under cooperative
agreement with the National Science Foundation.} and our own programs
written in C.
The signals from objects were recorded in two-dimensional (2D) CCD
detectors (called, ``frames" hereafter).
We carried out the reduction to obtain calibrated 2D spectra along
the spatial and wavelength axes with {\IRAF} as follows. 

{\bf Bias subtraction: } 
Even when the exposure time was zero, the pixel ADU counts were not 
zero but some positive value; a bias voltage was applied when reading 
the CCD to avoid negative values due to readout noise.
This bias value (ADU counts when the exposure time was zero) had to 
be subtracted from every frame.
We used an overscan region of 2D bias spectra to subtract the bias.

{\bf Flat fielding: }
We observed three dome flat fields for each grism and filter combination and
created high S/N dome flats by taking the median of these three dome flat frames.
We corrected for the non-uniformity by dividing the bias subtracted frame by
the median dome flat field.

{\bf Wavelength calibration: }
We used night glow OH emission lines recorded in object spectra for the 
red side and the Th-Ar lamp emission lines taken separately from object
frames for the blue side.
The overall wavelength calibration was done using a Legendre polynomial
of third order for the red side and a Legendre polynomial of fifth order for
the blue side. The polynomial order was determined to minimize the rms of
the wavelength residuals.

{\bf Distortion correction: }
Our spectra are distorted in the spatial direction to some extent.
We obtained a spatial distortion map for each frame
by tracing the spatial location of spectrophotometric standard stars at 
each wavelength.
We positioned the telescope so as to obtain spectra of objects and 
standard stars at the same position on the CCD.
We then used the distortion map to correct for each object spectrum.

{\bf Background subtraction: }
Background photons from night glow contaminated the frames.
We subtracted this background contamination by masking the
object region and fitting a low order polynomial function to the spatial
profile along the wavelength axis.

{\bf Flux calibration: }
We observed spectrophotometric standard stars: BD+28{\dosuu}4211, 
Feige110, GD71, BD+75{\dosuu}325 and Feige34. 
Library spectra in {\IRAF} were used for BD+28{\dosuu}4211, Feige110, 
BD+75{\dosuu}325 and Feige34. A spectrum in the HST Calibration Database 
System (CALSPEC) \footnote{\url{http://www.stsci.edu/hst/observatory/cdbs/calspec.html}} 
was used for GD71 to create the sensitivity function for the flux calibration, 
which is expressed as a fifth order cubic spline function. The best fit 
polynomial function was searched for each standard star spectrum. Table 
\ref{stdspec} is the list of spectrophotometric standard stars and the 
origin of the spectral energy distributions (SEDs) used in this reduction. 
The extinction file for the Mauna Kea telescope 
\footnote{http://www2.keck.hawaii.edu/inst/hires/mkoextinct.dat} was 
used to remove the telluric extinction.

{\bf Correction for dust extinction in our Galaxy: }
We corrected for dust extinction in our Galaxy using $R_V = 3.1$ 
and the color excess {\ebv} from \citet{sch98}. The extinction 
was assumed to follow the \citet{car89} law.

\begin{figure}
\epsscale{.80}
\plotone{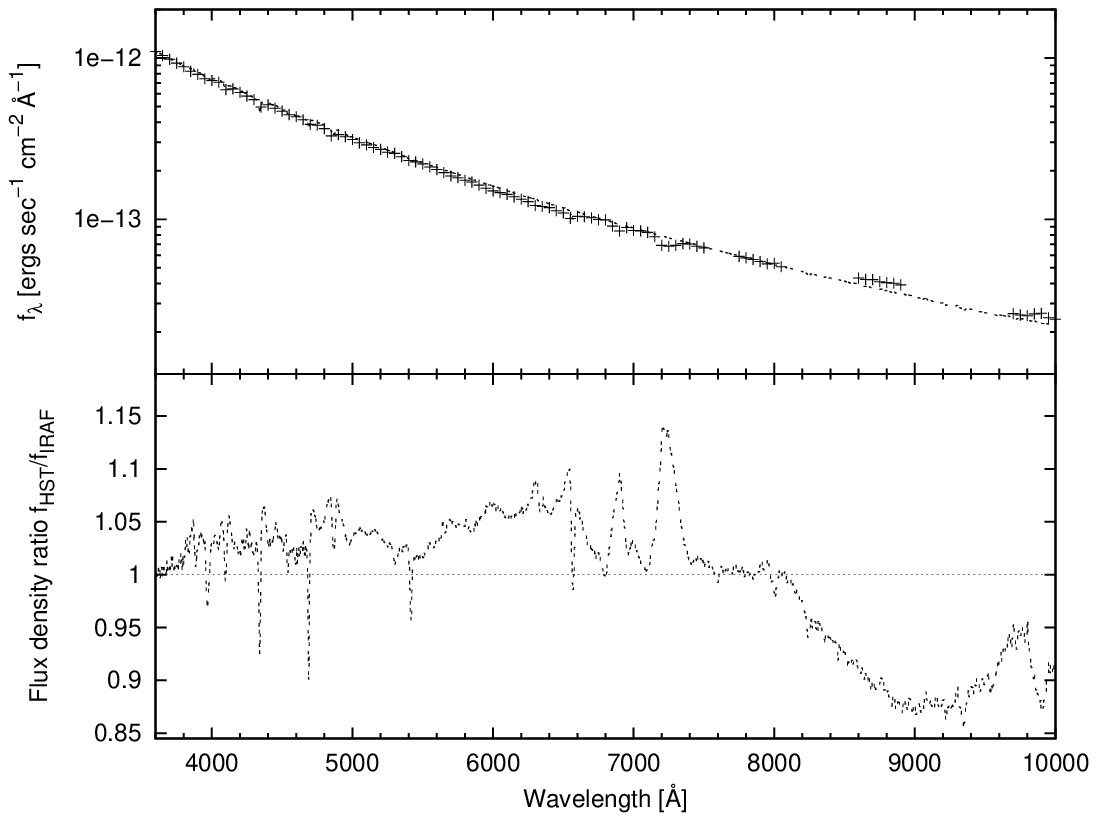}
\caption{Comparison of two BD+28{\dosuu}4211 spectra in the library. 
 Top panel: BD+28{\dosuu}4211 in the HST CALSPEC database
 \citep[line]{boh01} and in {\IRAF} \citep[plus]{mas88}. 
 Bottom panel: Flux ratio of the HST spectrum to the {\IRAF} spectrum.
 The IRAF spectrum is linearly interpolated to have flux densities
 at the same wavelength as the HST spectrum.
 These spectra differ by up to $10 \%$.\label{compbd}
}
\end{figure}

\begin{figure}
\epsscale{.80}
 \plotone{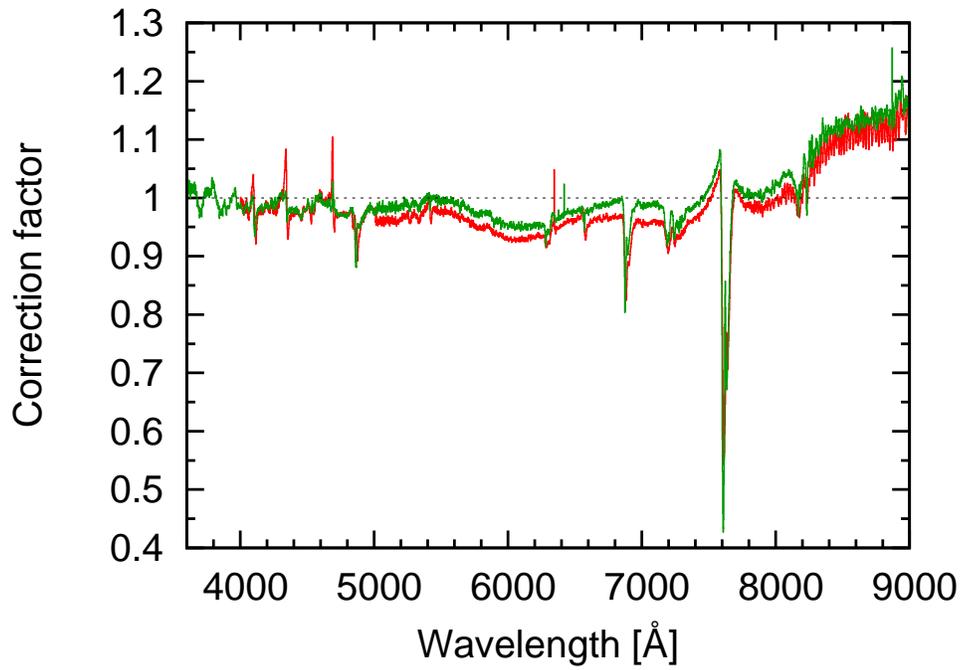}
 \caption{Correction factors for the telluric removal for year 2005 (red)
 and 2006 (green). There are calibration uncertainties of $<$4 \% near 
 6000 {\AA}, which are probably due to different telescope pointings 
 and atmospheric conditions.
 \label{telrm}}
\end{figure}

We developed C programs for the following processes:

{\bf Telluric absorption removal: }
Any spectra observed from the ground suffers from 
telluric absorption lines due to water and oxygen molecules in the 
atmosphere, especially at wavelengths longer than 5000 {\AA}.  The 
deepest is the A-band line at $\sim$7600 {\AA}.  For targets at $z \sim 0.25$, 
most of our sample, this absorption is superimposed around the ``{\sib}" 
or ``{\sic}" feature.  Since flux calibration is not perfect for the 
removal of sharp lines, we derived telluric-line correction 
factors which varies with wavelength using the following process:
(i) Composite spectra for each season (Year 2005 and 2006) were 
created from all the frames of a spectrophotometric standard star 
obtained during our campaign of the year.  BD+28{\dosuu}4211 was 
used for this purpose, since this was the best observed standard 
star during the relevant seasons (Table \ref{stdstar}).
(ii) A corresponding spectrum in the HST CALSPEC database was 
chosen for comparison. 
Library spectra of BD+28{\dosuu}4211 are also available in the 
{\IRAF} package; however, 
the size of the wavelength binning is smaller for the HST CALSPEC 
spectrum (5 {\AA}, black line in Figure \ref{compbd}), than the 
{\IRAF} spectrum (larger than 50 {\AA}, red pluses in 
Figure \ref{compbd}).  Moreover, the HST spectrum does not suffer 
from telluric absorption.  Note that up to a 10 \% difference in flux 
between these two spectra (bottom panel of Figure \ref{compbd})
is observed.
(iii) The extent of absorption at each wavelength was examined by 
comparison of these spectra.
Figure \ref{telrm} shows telluric-line correction factors for 
year 2005 (red) and 2006 (green). 
Calibration uncertainties are up to 4\% near 6000 {\AA} between these 
correction factors, perhaps due to different telescope pointing 
directions and atmospheric conditions at the observation times of the 
standard star and object observations.  This could produce a comparable uncertainty  
in the absolute flux calibration.                             . 
We performed the telluric-line correction by dividing an observed spectrum 
by correction factors from the same year.
It should be noted that residuals of telluric-lines could remain
to some extent after this process because of variable atmospheric conditions.

\begin{figure}
 \epsscale{.80}
 \plotone{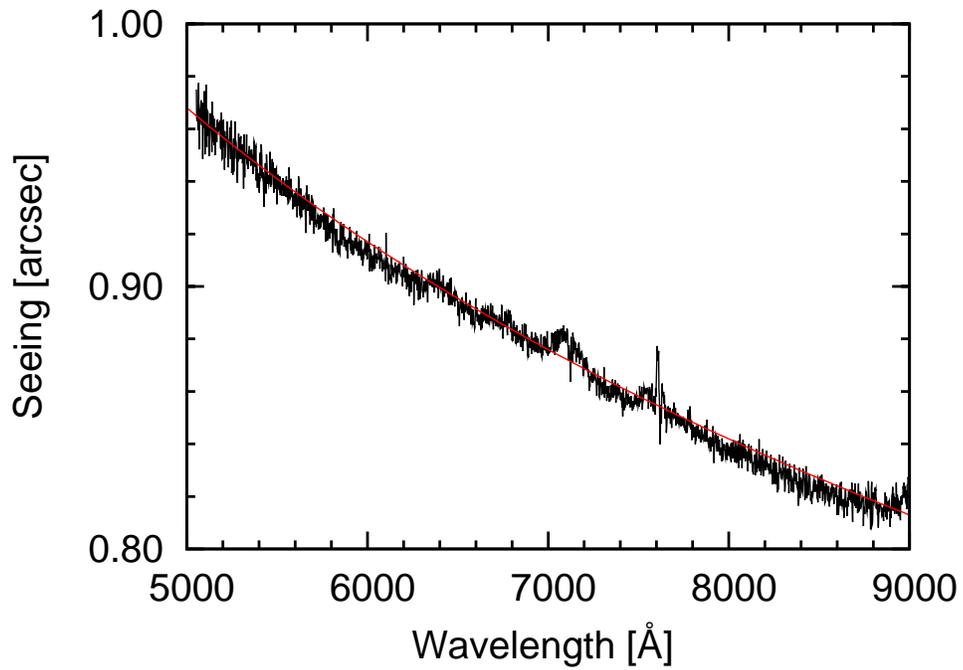}
 \caption{An example of the seeing-wavelength 
 relation measurement for the Subaru/FOCAS instrument. Plotted are 
 the seeings for BD+28{\dosuu}4211 taken on the night of Oct. 27, 2006. 
 The bestfit exponential function ($\theta \propto \lambda^{-0.3}$) is 
 shown in red. 
 \label{seeing}}
\end{figure}

{\bf Seeing measurement: }
Seeing is the Full Width at Half Maximum (FWHM) of the spatial 
profile recorded on the detector. 
We calculated the seeing using a 2D spectrum of a standard star, observed with 
a 2.0" slit (wide enough to pass through all the photons from a point-like object),  
along the spatial and wavelength axes.
The spatial profile of the star was traced in the wavelength 
direction with a Gaussian function to obtain the seeing size as a function of wavelength. 
We used BD+28{\dosuu}4211, since this was the best observed standard star. 
Figure \ref{seeing} is an example of the seeing size in arcsec 
measured for BD+28{\dosuu}4211 taken on the night of Oct. 27, 2006. 
The seeing is known to vary as an exponential with wavelength. 
We obtained the index of the seeing-wavelength 
relation for the Subaru/FOCAS instrument of $-0.297 \pm 0.001$:
\begin{equation}
 \theta \propto (\frac{\lambda}{\lambda_0})^{-0.3} \label{sw}.
\end{equation}
We used this relation for the SN extraction.
The index for the FOcal Reducer and low dispersion Spectrograph (FORS) attached 
to the Very Large Telescope (VLT) was reported to be -0.29 \citep{blo05}.
The Subaru/FOCAS result is comparable to their value. 

{\bf SN extraction: }
The observed SNe suffered from host galaxy contamination.
The best way to eliminate the galaxy light from the SN candidate may 
be to observe the same galaxy with the same telescope configuration
after the candidate disappears; however, this takes considerable time.
As an alternative, we have developed algorithms to eliminate the galaxy 
contribution.
Two proposed methods had been discussed in the literature:
one was to extract SN spectra from 1D galaxy-contaminated spectra 
with the aid of galaxy template spectra 
\citep[e.g.][]{how05,ell08,zhe08,fol08a,tok09,ost10} and the other was to
extract SN spectra from 2D galaxy-contaminated spectra without the
aid of galaxy template spectra \citep{blo05,bau08}.
An advantage of the latter method is that no assumption of a host galaxy
spectrum is necessary. For this method, one should determine the spatial 
profiles for the SN and host components for the extraction.
We developed an algorithm of SN extraction based on the latter concept.
The basic concept is similar to the technique used by the SuperNova
Legacy Survey collaboration \citep{bau08}, but developed independently. 

A model spectrum $M(x, \lambda)$ is assumed to be the sum of the SN and 
host galaxy flux:
\begin{equation}
 M(x, \lambda) = g_{SN}(x, \lambda) + g_{gal}(x, \lambda),
\end{equation}
where $x$ is the pixel index in the spatial direction and $\lambda$ is
the discretized wavelength.
The function $g_{SN}(x, \lambda)$ is the SN candidate flux density in the pixel
($x, \lambda$) and $g_{gal}(x, \lambda)$ is the host galaxy flux density. 
The spatial profile of a point source $g_{SN}(x, \lambda)$ is assumed to
be a Gaussian function:
\begin{equation}
 g_{SN}(x, \lambda) = h_{SN}(\lambda) 
  \exp \left(-\left( \frac{x-x_{SN}(\lambda)}{\theta_{SN}(\lambda)} \right) ^2 
  \right),
\end{equation}
where $\theta_{SN}$ is the bestfit Gaussian width for the SN candidate.
The spatial profile of the extended component $g_{gal}(x, \lambda)$ 
is assumed to be the sum of two Gaussian functions.  The summing was 
introduced in order to approximate the bulge and disk components of 
the host galaxy. 
We should note that a convolution with the PSF of more sophisticated 
functions such as the de Vaucouleurs $R^{1/4}$ law or an exponential 
law are more suited to fit galaxy morphologies.  Although a sum of 
two Gaussians is just an approximation, the $\chi^2$ of fitting 
tells us that this approximation is reasonable for the majority of our sample.  
For our hosts with complicated spatial profiles, we performed a 
$\chi^2$ fitting by adjusting two Gaussian peak positions representing 
the hosts.  A larger number of Gaussians or a more generalized function 
\citep{blo05} would be required for hosts with more complex spatial 
profiles.

In the fitting process, photon statistics was assumed to be
the only source of the uncertainty $\sigma(x, \lambda)$. 
This uncertainty was assumed to be constant in the spatial direction
$\sigma(x, \lambda) = \sigma(\lambda)$ and was estimated from the blank
areas adjacent to the object.
The width of the blank (no SN nor hosts) area was typically 25 arcsec on
both sides, centered on the object.  
We minimized:
\begin{equation}
 \chi^2(\lambda) = \sum_x 
  \left( \frac{O(x, \lambda) - M(x, \lambda)}{\sigma(\lambda)} \right)^2
\end{equation}
where $O(x, \lambda)$ was the observed flux density at each pixel.
 
The fitting was done with a two stage process.
First, the model function $M(x, \lambda)$ was fitted to the averaged
spatial profile created by averaging the 2D flux over the wavelength ranges
(3600 to 5400, 4000 to 6000 and 5000 to 9000 {\AA}) 
to determine the central positions and the widths of each component such as 
$\theta_{SN}$. The width $\theta_{SN}$ should be identical to the seeing size.
Then, the spatial profile at each wavelength was fitted, with
the spatial positions of the SN and its host fixed at the values obtained 
in the first step and
the widths changing due to the seeing-wavelength relation of Equation 
\ref{sw}. 

After finding the parameters, a candidate spectrum 
$g_{SN}(\lambda)$, its uncertainty $g_{SN,err}(\lambda)$ and the host 
galaxy spectrum $g_{gal}(\lambda)$ were derived.
If we created SN spectra by integrating flux over too wide a $x$ region, 
the S/N ratio became low.
We restricted SN candidate flux within the FWHM ($2 \sqrt{\ln 2} \theta_{SN}$).
\begin{align}
 g_{SN}(\lambda) = \int_{x_{smi} }^{x_{sma} } g(x, \lambda)dx, 
\end{align}
where $ x_{smi} \equiv x_{SN}-\theta_{SN} \sqrt{ln 2} \leq x \leq  
x_{SN}+\theta_{SN} \sqrt{ln 2} \equiv x_{sma}$, and the integral was
calculated using the Trapezoidal formula.

\begin{figure}[htbp]
 \begin{center}
  \leavevmode
  \subfigure[$r$-band image for slit alignment]% 
  {\includegraphics[width=7cm,height=5cm]
  {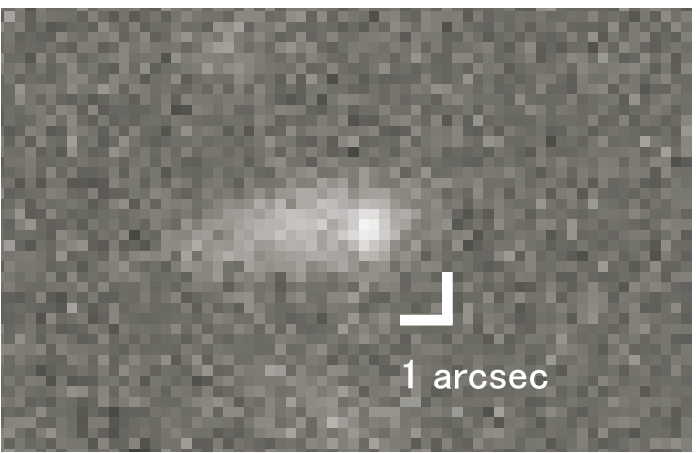}}
  \subfigure[2D spectrum in space (horizontal) and wavelength (vertical)]% 
  {\includegraphics[width=7cm,height=5cm]
  {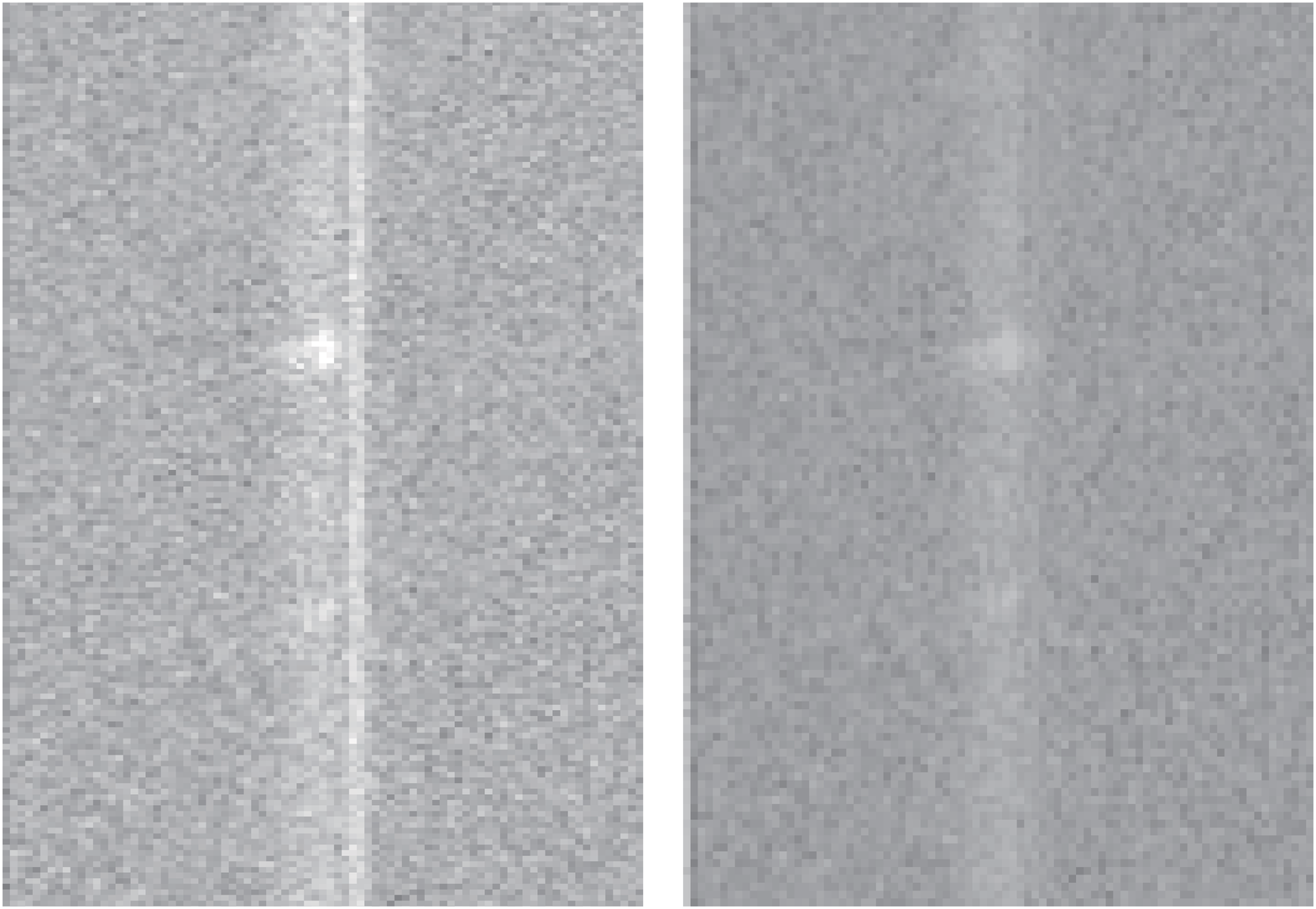}} \\
  \subfigure[Integrated spatial profile from a spectrum]%	
  {\includegraphics[width=7cm,height=5cm]
  {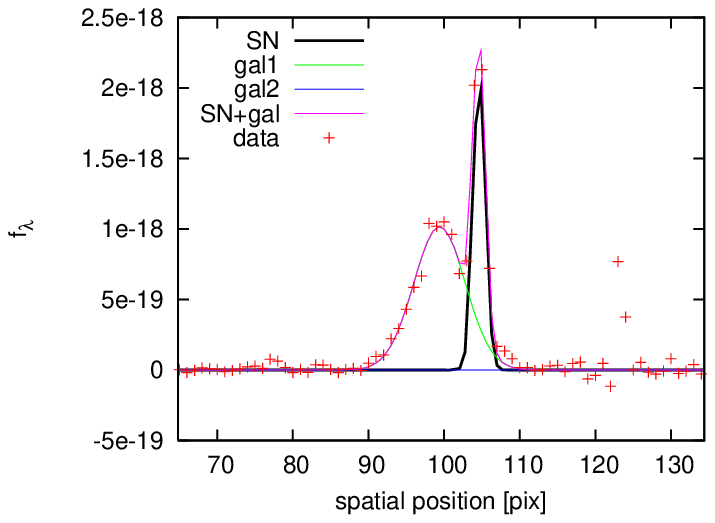}}
  \subfigure[Extracted SN spectrum]%	
  {\includegraphics[width=7cm,height=5cm]
  {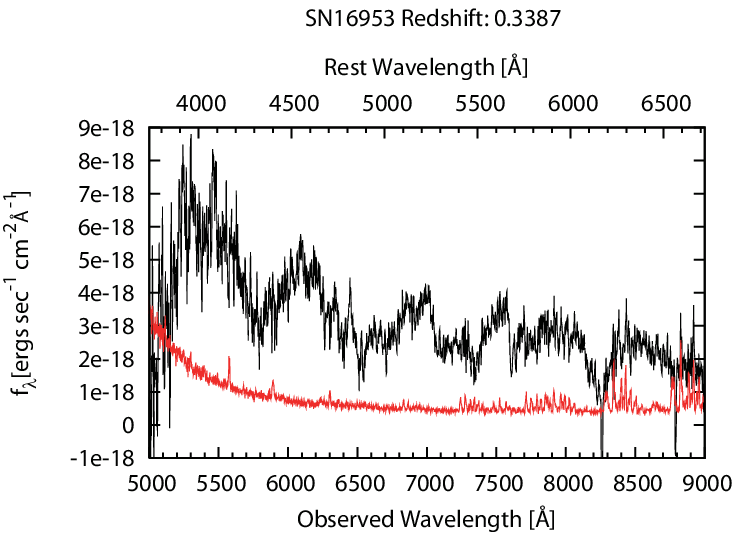}} \\
  
  \caption{An example of supernova spectrum extraction:
  (a) An $r$-band image; the SN Ia (SN16953) occurs on the right side of
  the host galaxy.
  (b) A part of the 2D spectrum centered on an emission line: before (left) 
  and after (right) SN extraction.
  (c) Spatial profile of the spectra: the observed profile (red) is fitted 
  with Gaussians for the SN (black) and galaxy (green) components. 
  The sum is shown with the pink curve.
  (d) Extracted SN spectrum (black) and uncertainty flux (red). 
  \label{sec}
  }
 \end{center}
\end{figure}

The SN flux uncertainty $g_{SN,err}(\lambda)$ was estimated from:
\begin{align}
 g_{SN,err}(\lambda) 
 &= \Delta (\int_{x_{smi}}^{x_{sma}} g_{SN}(x, \lambda)dx), \\
 \Delta h_{SN} &= \frac{\sigma(\lambda)}{\sqrt{\theta_{SN} \sqrt{\pi/2}}}.
\end{align}
Therefore, 
\begin{align}
 g_{SN,err}(\lambda) 
 &= \int_{x_{smi}}^{x_{sma}} 
 \exp \left( - \left( \frac{x-x_{SN}}{\theta_{SN}} \right)^2 \right) dx
 \times \Delta h_{SN} \theta_{SN}, \nonumber \\ 
 &=  \int_{x_{smi}}^{x_{sma}} 
 \exp \left( - \left( \frac{x-x_{SN}}{\theta_{SN}} \right)^2 \right) dx \nonumber \\ 
 & \times \left(\frac{2}{\pi} \right)^{1/4} \sqrt{\theta_{SN}} \sigma(\lambda). 
\end{align}
Host galaxy spectra were constructed by subtracting the SN component
$g_{SN}$ from $O(x, \lambda)$.
In Figure \ref{sec}, we show an example of the SN extraction procedure.

We should note that 
the fitting would be affected by the presence of emission lines.
Our method fits spatial profiles at each wavelength with the sum of two
or three Gaussian functions. When a host galaxy spatial profile can be 
approximated by a Gaussian, the total spatial profile would be the sum of
two Gaussians. For the galaxy approximated by two Gaussians, the total would
be that of three Gaussians.
Spatial profiles at the lines could be too complicated for our model if there were
strong lines from host galaxies. 
Since SNe Ia do not show emission lines in their spectra, we can regard emission lines
in the extracted SN spectrum, more than 80 \% of our sample, as those originated in 
their host galaxies. We can regard absorption lines in these spectra as galaxy origin
if its width is smaller than expanding velocities of SN ejecta (\vel{$\sim 10,000$}).

A profile fit to a Gaussian function might miss a few percent of the 
observed flux. An analysis of spatial profiles for standard stars
showed a slight excess in flux at the outskirts of the spatial profile
compared to the prediction of the Gaussian function. 
We neglect this excess for our current analysis.

{\bf Combining spectra: }
In most cases, SN spectra were taken with two different combinations of
grisms and filters: blue and red side spectra.
We combined these two spectra by taking the average flux within the
overlapping wavelength regions and scaling the red side to match the blue
side flux and re-binning to 2 {\AA} per pixel.
The result is a contiguous science spectrum together with an uncertainty
spectrum representing the statistical uncertainties in the flux in
each binned pixel.

{\bf Slitloss correction: }
Since seeing sizes were comparable to the slit width, some fraction of the object photons could not go through the slit. We corrected for this slitloss by estimating $gri$ magnitudes at the spectroscopy date from light curves and spectra. 
Magnitudes at the spectroscopy date were determined by interpolating the bestfit templates for the SALT2 light curve fitter (see the light curve fitting in \S \ref{salt2fit}). 
Magnitudes were synthesized from spectra using $gri$ filter responses.
We matched the interpolated magnitudes with the synthetic magnitudes to determine a second-order polynomial slitloss correction function; 
\begin{equation}
 a(\lambda) = \sum_{i=0}^2 c_i \lambda^i, \label{specphoto}
\end{equation} 
where $c_i$ is a coefficient. This function was determined for each spectrum.

\section{Dataset} \label{datasets}
\S \ref{zz} describes the method for determining types and
redshifts for all the Subaru targets and the following sections 
are dedicated only to SNe Ia.

\subsection{Determining redshifts and supernova types \label{zz}}

\begin{figure}
 \epsscale{1.0}
 \plotone{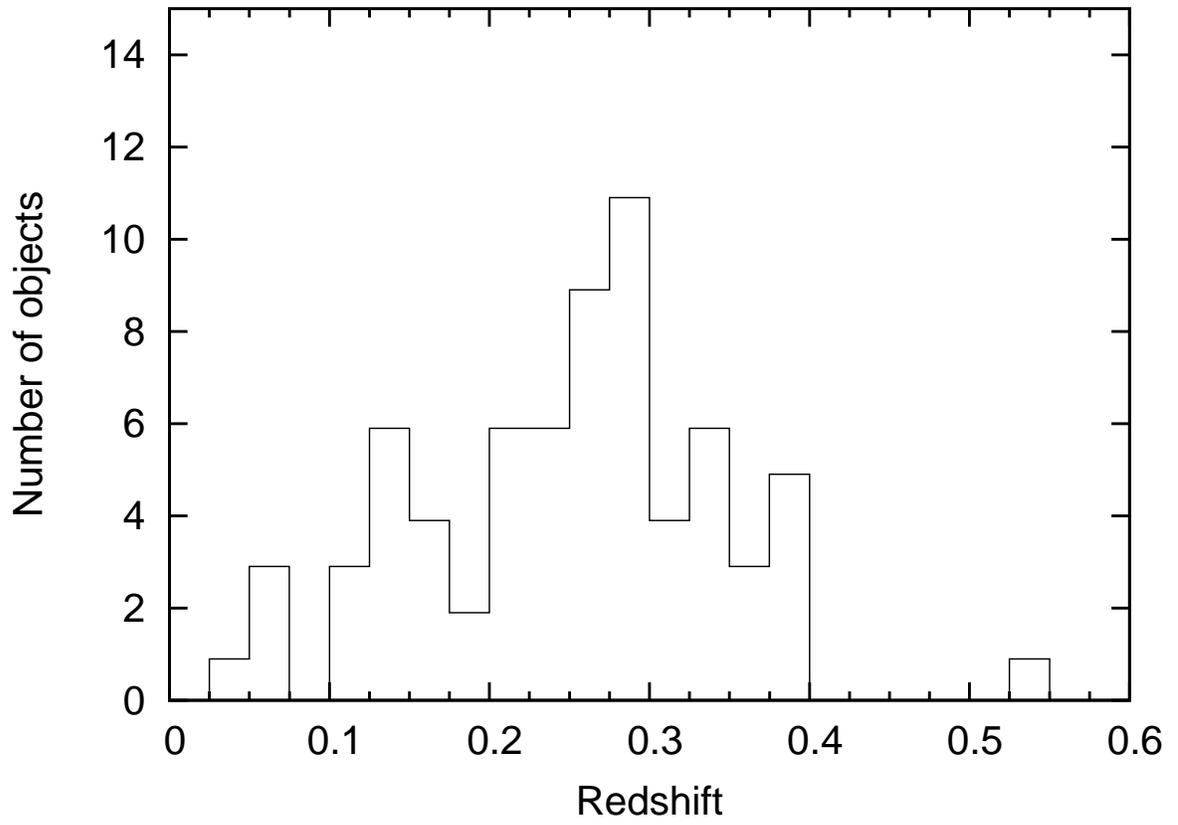}
 \caption{Redshift distribution of the 71 observed objects with the Subaru telescope.
 \label{sbr_z}
 }
\end{figure}

One can determine the redshift of the supernova either from the 
supernova itself or from the host galaxy; the latter was our 
primary choice.  The representative emission lines are Balmer 
lines ({\ha} and {\hb}) and forbidden lines of [{\NII}], [{\OII}], 
[{\OIII}] and [{\SII}].
Redshifts were determined from the wavelengths determined by fitting Gaussian functions to all the emission lines. The absorption lines we used were Ca H\&K, Mgb (SN12991) and NaD (SN12979). We identified absorption lines by visual inspection. The final redshift was the average value of all measured lines with equal weight. Lines and their rest-wavelengths \footnote{http://www.sdss.org/dr7/algorithms/speclinefits.html} are listed in Table \ref{emit}.
If there were no spectral lines in the host galaxy spectra, we adapted the technique of spectra template fitting to the SN spectrum. This technique cross-correlates SN spectra with a series of SN library spectra (spectral templates and high quality nearby SN spectra) and search for the $\chi^2$ minimum to determine redshifts and types; details are explained in \citet{tok09}.
The redshifts for 64 objects were determined from spectral lines of their host galaxies and seven were from the fitting of SN spectra. Figure \ref{sbr_z} shows the redshift distribution for all our targets except 1 AGN at $z = 1.3716$.

We assigned SN classification by visual inspection \citep{fil97}: 
SNe Ia are characterized by a deep absorption trough at $\sim 6100$ 
{\AA} produced by a blue-shift of ``{\sic}''.  SNe Ib show deep 
absorption of blueshifted {\HeI} at {\lamlam 4471,5876,6678} at 
similar phases.  SNe Ic show several strong {\CaII} and {\OI} 
{\lam 7774} absorption lines along with no hydrogen lines and a 
very weak ``{\sic}" line.
SNe II all prominently exhibit hydrogen in their spectra.  AGN show 
very broad emission lines of hydrogen Balmer lines and forbidden 
lines on a power-law continuum.  For the seven spectra not 
classified visually due to low S/N, a spectral template fitting 
\citep{tok09} was used.  Our type classifications resulted in 62 
confirmed SNe Ia, 3 of which have not been assigned IAU names, 
8 SNe II, and 1 AGN.

Table \ref{sbrspecinfo} describes SN candidates observed by the Subaru/FOCAS telescope.
Column 1 is the SDSS identification number.
We denote each SN by its SDSS identification number in this 
paper, but show its IAU name in Column 2.  Column 3 is the candidate type.
The celestial coordinates of candidates are shown in Columns 4 and 5.
Column 6 is the color excess for our Galaxy from the \citet{sch98} map.
Column 7 is the mean time of our observation. 
When both blue and red side spectra were taken, which is the case for most of our sample, we defined the observation date as the average time of both observations.
Column 8 is the redshift. The parenthesis in this column represents how the redshift was determined; ``ge'' from galaxy emission lines, ``ga'' from galaxy absorption lines, ``sn'' from the SN
spectrum fitting and ``snh'' from the {\ha} line of SNe II.
Column 9 is the mean airmass of our observations.  The median value is 1.2, 
with 23 \% of the observations having an airmass larger than 1.3. 
Column 10 shows two ratios, $F_{host}/F_{all}$, of host galaxy flux to reconstructed 
SN and host flux for the blue (4000 to 6000 {\AA} or 3600 to 5400 {\AA}) and red 
side spectra (5000 to 9000 {\AA}), respectively.  Ratios are determined from 
averaged spatial profiles over the wavelength region of the spectrum, which 
was used to determine the central position and width of each component in 
the SN extraction.  The spatial region was integrated over the FWHM centered 
on the SN position.  When either the blue or red side is missing, this is  
represented in the table by "--".  When a SN occurs in a host core, SN and 
host profiles can be degenerate.  This is especially the case when the 
brightness of the host is comparable to the SN.  Since these candidates can 
be a variable active galactic nucleus, a lower priority was put for these SN 
candidates for spectroscopic followup \citep{sak08}, and thus we have few 
such candidates.  For 62 of 66 SN candidates with blue and red side spectra, 
the difference of host contamination between the two sides 
($|F_{host,blue}-F_{host,red}|$) was less than 0.2.  The difference was 
larger than 0.4 for the remaining four candidates (SN5751, 16779, 16938, 
and 17081).  Either side of the first three (SN5751, 16779 and 16938) shows 
a degenerate spatial profile and have large host contamination (around 0.6).  
The bestfit Gaussian width $\theta_{SN}$ for the spatial profile of these 
SNe was larger than the width of their host.  Our SN extraction method did 
not work sufficiently.  A different subtraction method based on the 
principle component analysis is adapted for these 
spectra (Sako et al. in prep).  The Gaussian width for SN17081 was normal 
($\theta_{SN}=1.2$), corresponding to the FWHM of $0.6''$.  From its 
negative value of $F_{host,blue}-F_{host,red}$ (-0.40), the lowest 
contamination ratio of these four candidates, we speculate that SN17081 
occurred in a red galaxy. 

\begin{figure}
 \plotone{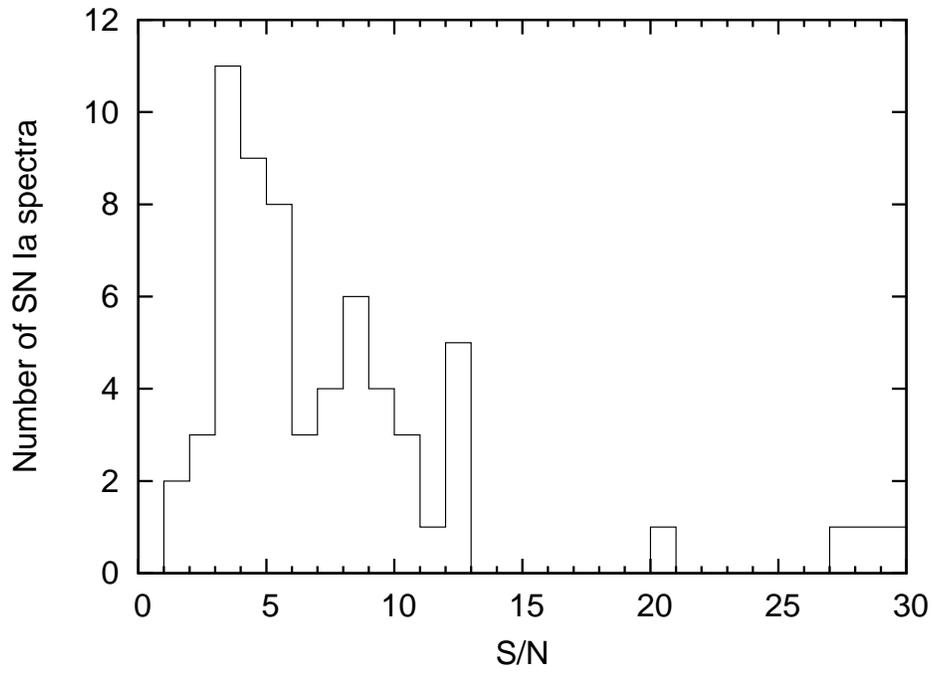}
 \caption{Distribution of the S/N for the SN Ia spectra. The S/N is  calculated per 2 {\AA}.
 \label{sn_sn1a}}
\end{figure}

\begin{figure}[htbp]
 \begin{center}
  \leavevmode
  \subfigure[$r$-band image for slit alignment]% 
  {\includegraphics[width=7cm,height=5cm]
  {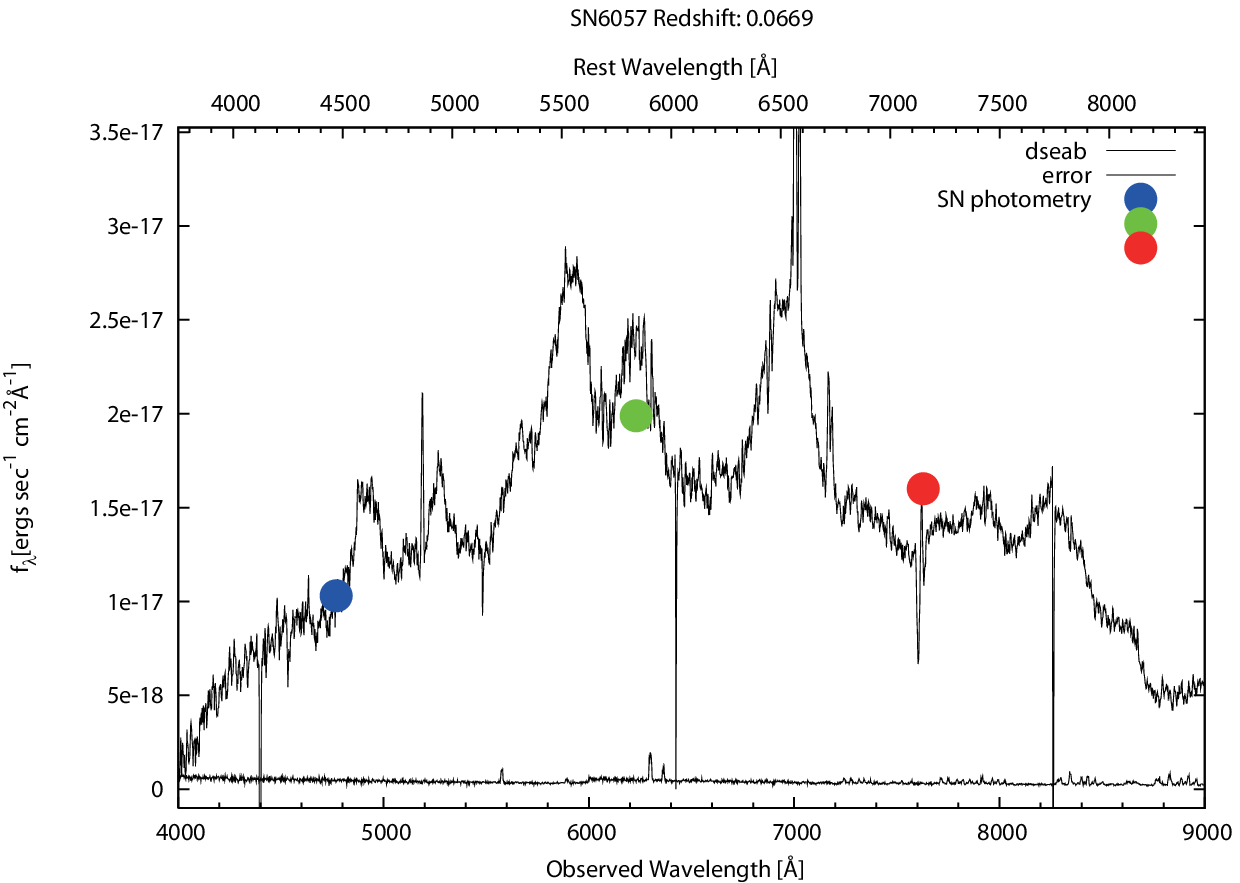}}
  \subfigure[2D spectrum in space (horizontal) and wavelength (vertical)]% 
  {\includegraphics[width=7cm,height=5cm]
  {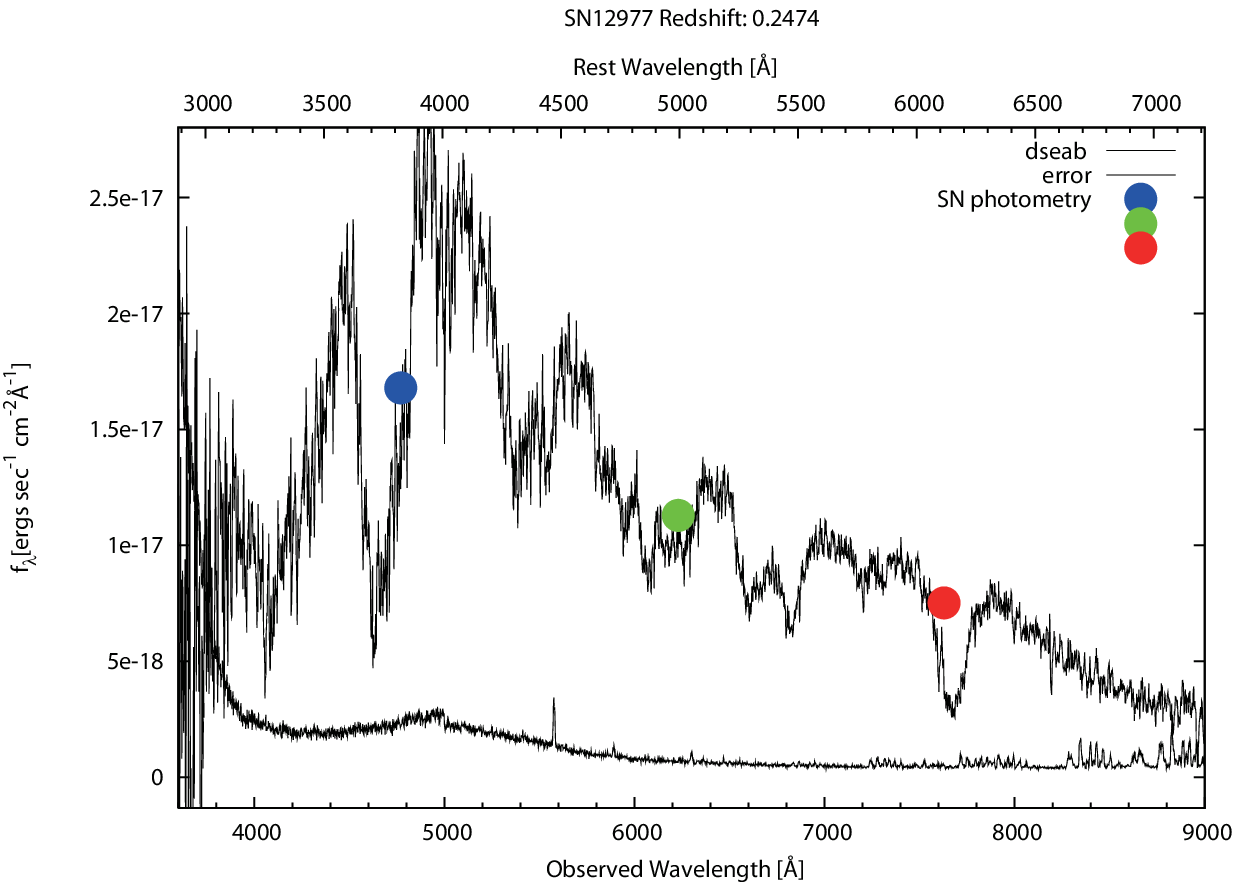}} \\
  \subfigure[Integrated spatial profile from a spectrum]%	
  {\includegraphics[width=7cm,height=5cm]
  {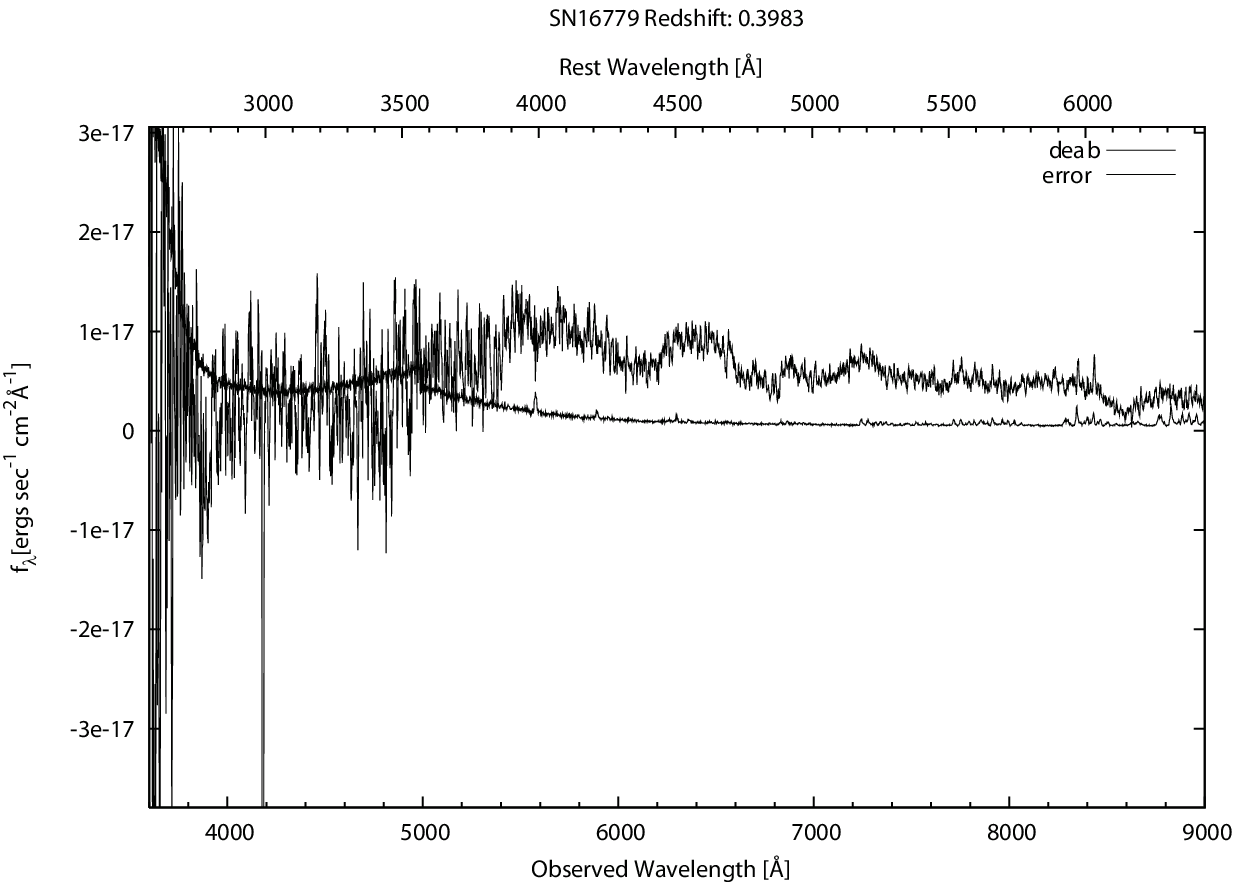}}
  \subfigure[Extracted SN spectrum]%	
  {\includegraphics[width=7cm,height=5cm]
  {f6d.eps}} \\
  
  \caption{Examples of SN Ia spectra and their uncertainties obtained by the Subaru/FOCAS instrument: two high quality SN Ia spectra: late-time (one month after the maximum date) spectrum of a nearby SN Ia $z=0.0669$ in the top left, a near maximum spectrum of mid-z SN Ia at $z=0.2474$ in the top right, and one low quality SN Ia spectrum (one week after the maximum date) at $z=0.3983$ in the bottom left. The lower and upper horizontal axes are observed and rest wavelengths. The vertical axis is the flux density. Spectra are corrected for slitloss if the $gri$ magnitudes can be interpolated at the spectral date (labeled as ``dseab" in the figures; $g$, $r$ and $i$ magnitudes are shown in blue, green and red colors) or not (labeled as ``deab" in the figure). \label{spec_ex}
  }
 \end{center}
\end{figure}

For each SN, we measured the average S/N over the 
entire spectral range in 2 {\AA} bins.  Figure \ref{sn_sn1a} shows 
the S/N distribution for the 62 SNe Ia.  Most spectra of the sample 
show S/N between 4 and 10.  The very high S/N spectra are those of 
SNe Ia at $z \sim 0.05$.  Figure \ref{spec_ex} gives examples of 
SN Ia spectra obtained by the Subaru/FOCAS telescope: two high 
quality SN Ia spectra are shown, a spectrum of a nearby SN Ia at $z=0.0669$ 
one month after from the maximum date in the top left, 
and a near maximum spectrum of mid-z SN Ia at $z=0.1404$ in the top right.  
Also a SN Ia spectrum at $z=0.3983$ of low quality one week after 
from the maximum date is given in the bottom left.  

The Subaru telescope observed 71 SN candidates in total with the FOCAS 
instrument and confirmed 59 normal SNe Ia and three probable SNe Ia. 
Table \ref{obsf} summarizes the observational results of the 
Subaru/FOCAS SDSS SN campaign.
The template fitting of Subaru spectra has provided the bestfit SN 
type (and redshifts in case of no lines from their hosts) for 
ambiguous targets based on the reduced $\chi^2$.  The following are 
notes on those targets. 
(i) SN6471. The bestfit type among our template spectra is a SN IIP, 
matching the SN1999em and SN1992H spectra well.  We could not detect 
emission or absorption lines of this host.  The {\ha} emission of 
SN6471 shows that SN6471 is one of the farthest SNe IIP 
discovered by the SDSS-II Survey, $z=0.202$. 
(ii) SN10450. The {\hb} and [{\OII}] emission lines of this host 
show that this is one of the farthest SNe discovered by the 
SDSS-II SN Survey, $z=0.5399$.  Our library does not contain 
spectra that match all the spectral features of SN10450. An 
emission line around 7500 {\AA} in the observer frame may be 
broader than nebular emission.  If this is the case, SN10450 is 
a peculiar SNIIn.  SN10450 is around 1 mag brighter in the rest 
$r$ band than normal SNe Ia at this redshift. 
(iii) SN12844. The bestfit type among our template spectra is 
SN2005gj, which shows an {\ha} feature evolving with time and called a 
Type Ia/IIn SN \citep{ald06,pri07}. The spectrum of SN12844 shows strong emission
lines of [{\OII}], [{\OIII}], {\hb}, {\ha}, [{\NII}] and [{\SII}].
(iv) SN12979. The bestfit type among our template spectra is 
an intrinsically faint SN1991bg.
(v) SN14475. The bestfit spectrum is SN1998bw \citep{gal98,pat01} at $+6$ days 
after the maximum date.  Other matches are SN1997ef \citep{iwa00} and SN2002ap 
\citep{gal02,maz02,fol03}. 
Although all the details of the absorption lines of the bestfit spectra are 
not completely identical, they are all hypernovae.  Thus it can be said that 
SN14475 is a possible hypernova.
(vi) SN15170. The bestfit spectrum is the intrinsically bright SN1991T spectrum around 
the maximum date. 

\subsection{Lightcurve properties \label{salt2fit}}

We used the SALT2 light curve fitting code developed by the SNLS collaboration \citep{guy07} to derive light curve parameters. This code employs a two dimensional surface in time and wavelength that describes the temporal evolution of a SN Ia spectrum $f(p, \lambda)$;
\begin{equation}
 f(p, \lambda) = x_0 \times [M_0(p, \lambda)+x_1 M_1(p, \lambda)]
  \times \exp(c_{salt} CL(\lambda)),
\end{equation}
where $p$ is the SN phase in restframe days, the elapsed time from maximum luminosity, $M_0$ is the average SED, and $x_0$ is a normalization. 
The first-order spectral deviation $M_1$ is included in a linear fashion 
with a coefficient {\xx}.  We used the spectral surfaces of $M_i$ 
($i=0,1$) that we also used for the cosmological analysis of \citet{kes09}. 
The parameter {\xx} is the normalized deviation from the typical spectrum $M_0$ and can be approximated with {\dm} as $\sim 1.1 - 0.16${\xx} \citep{guy07}.
The function $CL(\lambda)$ is the color law of a third order polynomial function monotonically increasing with wavelength and {\csalt}, the deviation from the mean SN Ia (B-V) color, is its coefficient which is sensitive to both intrinsic color diversity and the host-dust reddening. 

We selected every photometric point of the $gri$ band light curves unless a bad $flag$ was assigned; 
this happened when the photometric scaling factor was low, the rms magnitude of calibration stars was high, the variation in sky brightness was high, the frame fit quality for the SN photometry was poor, or no calibration stars were available in the frame (corresponding to the $flag < 256$ as recommended in \citet{hol08}). We did not use $u$ and $z$ band data, since their S/N were not good enough for detailed analysis.

\begin{figure}
 \epsscale{1.0}
 \plotone{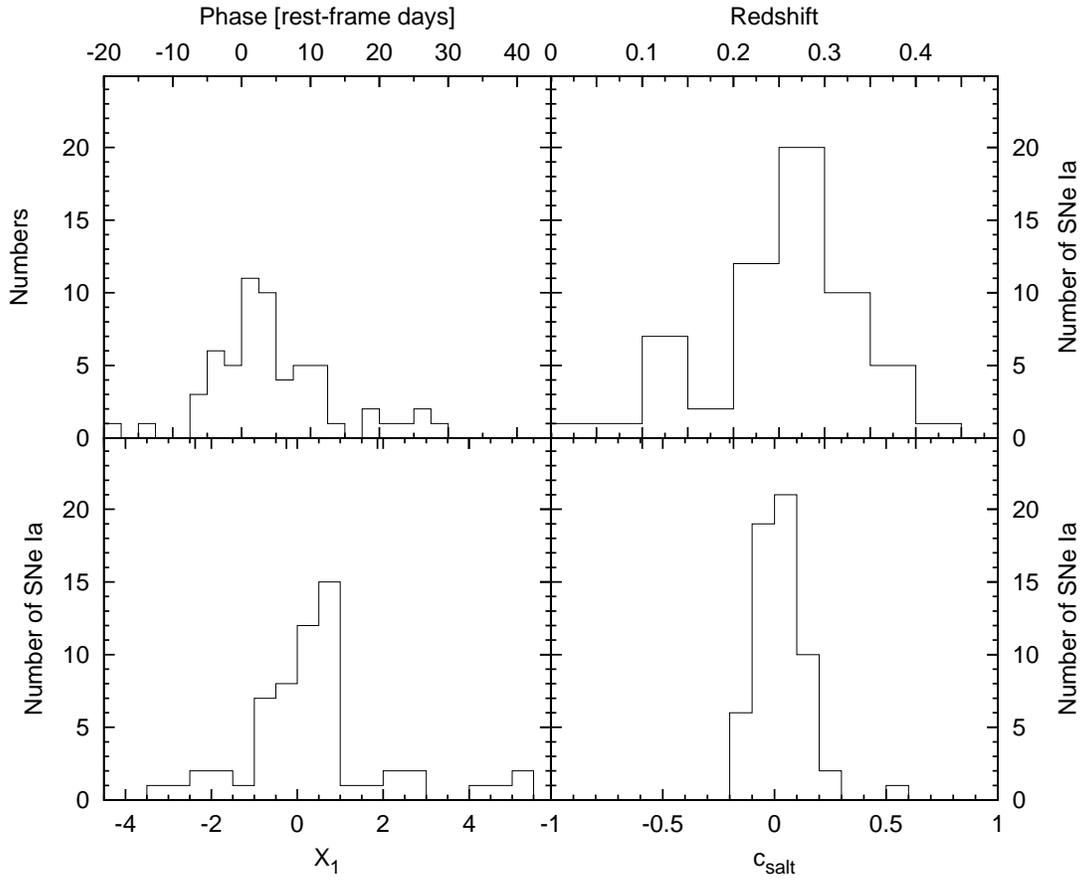}
 \caption{SN phase (top left), redshift (top right), spectral deviation {\xx} (bottom left) and color {\csalt} (bottom right) distributions for the Subaru sample. 
 \label{sbrdist}
 }
\end{figure}

The output parameters of this code were rest-$B$ band maximum magnitude $m_B$ 
, which is derived from $x_0$, spectral deviation {\xx}, 
color excess {\csalt}, and the date of maximum luminosity {\tmax}. 
In Figure \ref{sbrdist}, we present the distribution of SN phases at the spectroscopic epoch (upper left), the redshift (upper right), the spectral deviation {\xx} (lower left) and the color excess {\csalt} (lower right) for the Subaru sample. 
The SN phase is defined as the difference of dates between the 
spectral observation and maximum brightness divided by $(1+z)$. We can 
see that the range of phases varies from -7 days to +30 days and that 
the highest number is in the bin $p=0-5$ after maximum brightness (top 
left panel).
The redshift distribution reaches to the $z=0.40-0.45$ bin and it 
peaks at the $z=0.25-0.30$ bin (top right).
The lightcurve-width histogram resembles a right rectangle shape, with 
its oblique side decreasing with the value of {\xx}.  The number 
shortage of SNe Ia with large {\xx} values is due to a bias that arose 
when the Subaru targets were selected (left bottom). 
Most of the sample fall in the color range from -0.2 to 0.2.  The SALT2 
fitting gives large positive values of this parameter for ambiguous 
targets SN12844, SN12979 and SN14475, which are {\csalt} 
$= 0.66 \pm 0.04$, $0.91 \pm 0.06$, and $0.71 \pm 0.03$, respectively. 
These might be cool or dusty SNe. 

\section{Summary} \label{summary}
We carried out followup spectroscopy of 71 SN Ia candidates discovered by the SDSS-II Supernova Survey. Nearly 90 \% of the observed spectra were identified as SNe Ia, showing the high efficiency of the photometric selection. SN cosmology and related work, such as the study of spectroscopic diversity, will benefit from our observations. The number of SNe Ia in the intermediate redshift range has increased substantially. 
Although observed spectra are a combination of the SN flux and the flux of their host galaxies, we succeeded in extracting the SN component from the observed spectra using code developed by the author. 

\acknowledgements
Acknowledgements -- 
We thank Harold Spinka for his careful reading of the manuscript for publication. K.K. thanks the COE Program ``the Quantum Extreme Systems and Their Symmetries'' for fiscal 2007, the Global COE Program ``the Physical Sciences Frontier'' for fiscal 2008-2009, MEXT, Japan and the JASSO scholarship for fiscal 2007-2009. L.{\"{O}} is partially supported by the Spanish Ministry of Science and Innovation (MICINN) through the Consolider Ingenio-2010 program, under project CSD2007-00060 ``Physics of the Accelerating Universe (PAU)''.

Funding for the SDSS and SDSS-II was provided by the Alfred P. Sloan Foundation, the Participating Institutions, the National Science Foundation, the U.S. Department of Energy, the National Aeronautics and Space Administration, the Japanese Monbukagakusho, the Max Planck Society, and the Higher Education Funding Council for England. \url{The SDSS Web site is http://www.sdss.org/}.

The SDSS is managed by the Astrophysical Research Consortium (ARC) for the Participating Institutions. The Participating Institutions are The University of Chicago, Fermilab, the Institute for Advanced Study, the Japan Participation Group, The Johns Hopkins University, Los Alamos National Laboratory, the Max-Planck-Institute for Astronomy (MPIA), the Max-Planck-Institute for Astrophysics (MPA), New Mexico State University, University of Pittsburgh, Princeton University, the United States Naval Observatory, and the University of Washington. This research has made use of the NASA/IPAC Extragalactic Database (NED) which is operated by the Jet Propulsion Laboratory, California Institute of Technology, under contract with the National Aeronautics and Space Administration. 

Facilities: \facility{SDSS}, \facility{Subaru(FOCAS)}

\clearpage

\begin{deluxetable}{cccc}
 \tabletypesize{\scriptsize}
 \tablecaption{Instrument Configurations \label{setups}} 
 \tablewidth{0pt}
 \tablehead{
 \colhead{Side} & \colhead{Grism} & \colhead{Filter} & 
 \colhead{Spectral Region {\AA}}
 }
 \startdata
 2005 & & & \\
 \hline
 blue & 300B & L600 & 4000-6000 \\
 red  & 300R & SY47 & 5000-9000 \\
 \hline
 2006 & & & \\
 \hline
 blue & 300B & L550 & 3600-5400 \\
 red  & 300R & SY47 & 5000-9000
 \enddata
 \tablecomments{Detailed information is found at: 
 http://www.naoj.org/Observing/Instruments/FOCAS/spec/grisms.html}
\end{deluxetable}
\clearpage

\begin{deluxetable}{lccc}
%\tabletypesize{\scriptsize}
%\rotate
\tablecaption{The spectrophotometric standard stars \label{stdstar}}
\tablewidth{0pt}
\tablehead{
 \colhead{Name (grism \& filter)} &  \colhead{Date} & \colhead{ZD (deg)} & 
 \colhead{Exposure (s)} 
}
\startdata
\hline
2005 \\
\hline 
BD+28{\dosuu}4211(300B \& Y47) & Sep26 & 35.0 & 10.0 \\ 
BD+28{\dosuu}4211(300R \& L600) & Sep26 & 34.7 & 20.0 \\ 
Feige110(300B \& Y47) & Sep26 & 31.4 & 30.0 \\ 
Feige110(300R \& L600) & Sep26 & 31.9 & 40.0 \\ 
BD+28{\dosuu}4211(300B \& Y47) & Sep27 & 39.0 & 20.0 \\ 
BD+28{\dosuu}4211(300R \& L600) & Sep27 & 38.2 & 10.0 \\ 
GD71(300B \& Y47) & Sep27 & 5.8 & 90.0 \\ 
GD71(300R \& L600) & Sep27 & 5.3 & 90.0 \\ 
BD+28{\dosuu}4211(300B \& Y47) & Oct25 & 17.8 & 15.0 \\ 
BD+28{\dosuu}4211(300R \& L600) & Oct25 & 16.7 & 10.0 \\ 
BD+75{\dosuu}325(300B \& Y47) & Oct25 & 55.4 & 5.0 \\ 
BD+75{\dosuu}325(300B \& Y47) & Oct25 & 55.4 & 5.0 \\ 
BD+28{\dosuu}4211(300B \& Y47) & Oct26 & 18.1 & 15.0 \\ 
BD+28{\dosuu}4211(300R \& L600) & Oct26 & 17.8 & 10.0 \\ 
BD+75{\dosuu}325(300B \& Y47) & Oct26 & 55.3 & 4.0 \\ 
BD+75{\dosuu}325$^a$(300R \& L600) & Oct26 & 55.3 & 4.0 \\ 
BD+28{\dosuu}4211(300B \& Y47) & Oct27 & 16.8 & 30.0 \\ 
BD+28{\dosuu}4211(300R \& L600) & Oct27 & 16.1 & 20.0 \\ 
GD71(300B \& Y47) & Oct27 & 25.0 & 60.0 \\ 
GD71(300R \& L600) & Oct27 & 25.6 & 60.0 \\ 
Feige110(300B \& Y47) & Nov25 & 26.4 & 60.0 \\ 
Feige110(300R \& L600) & Nov25 & 25.9 & 40.0 \\ 
Feige34(300B \& Y47) & Nov25 & 27.3 & 15.0 \\ 
Feige34(300R \& L600) & Nov25 & 27.6 & 25.0 \\ 
Feige110(300B \& Y47) & Nov26 & 26.6 & 60.0 \\ 
Feige110(300R \& L600) & Nov26 & 26.4 & 40.0 \\
\hline
2006 \\
\hline 
BD+28{\dosuu}4211(300B \& L550) & Sep15 & 36.0 & 3.0 \\ 
BD+28{\dosuu}4211(300B \& L550) & Sep16 & 43.3 & 15.0 \\ 
BD+28{\dosuu}4211(300R \& Y47) & Sep16 & 41.3 & 15.0 \\ 
BD+28{\dosuu}4211(300B \& L600) & Sep16 & 40.7 & 10.0 \\ 
BD+28{\dosuu}4211(300B \& L550) & Sep17 & 45.8 & 30.0 \\ 
BD+28{\dosuu}4211(300B \& L600) & Sep17 & 41.3 & 10.0 \\ 
BD+28{\dosuu}4211(300R \& Y47) & Sep17 & 42.0 & 15.0 \\ 
GD71(300B \& L550) & Sep17 & 16.8 & 20.0 \\ 
GD71(300B \& L600) & Sep17 & 14.9 & 10.0 \\ 
GD71(300R \& Y47) & Sep17 & 15.3 & 15.0 \\ 
BD+28{\dosuu}4211(300B \& L550) & Oct14 & 23.4 & 10.0 \\ 
BD+28{\dosuu}4211(300R \& Y47) & Oct14 & 23.7 & 15.0 \\ 
BD+28{\dosuu}4211(300B \& L550) & Nov25 & 12.9 & 5.0 \\ 
BD+28{\dosuu}4211(300R \& Y47) & Nov25 & 13.1 & 15.0 \\ 
Feige34(300B \& L550) & Nov25 & 29.5 & 40.0 \\ 
Feige34(300R \& Y47) & Nov25 & 29.2 & 60.0 \\ 
Feige110(300B \& L550) & Nov26 & 35.2 & 15.0 \\ 
Feige110(300R \& Y47) & Nov26 & 35.5 & 30.0 \\ 
\enddata
%\tablecomments{}
%\tablenotetext{a}{}
\end{deluxetable}

\begin{deluxetable}{lll}
 \tablecaption{Library spectroscopic standard stars used in this study 
 \label{stdspec}}
 \tablewidth{0pt}
 \tablehead{
 \colhead{Name} & \colhead{Location of file} & \colhead{Ref.}
 }
 \startdata
 BD+28{\dosuu}4211 & spec50cal \tablenotemark{a} & \citet{mas88} \\
 Feige 110   & spec50cal \tablenotemark{a} & \citet{mas88} \\
 GD 71       & calspec   \tablenotemark{b} & \citet{boh01} \\
 BD+75{\dosuu}325    & oke1990   \tablenotemark{c} & \citet{oke90} \\
 Feige 34    & spec50cal \tablenotemark{a} & \citet{mas88} \\
 \enddata
 \tablenotetext{a}{This directory is under \$onedstd/ in {\IRAF}}
 \tablenotetext{b}{Downloaded from the Calibration data system;
 \url{http://www.stsci.edu/hst/observatory/cdbs/calspec.html}.
 Rebinned to 50 {\AA}.}
 \tablenotetext{c}{This directory is under \$onedstd/ in 
 {\IRAF}. We used this spectrum after a magnitude correction by
 +0.004mag. \citep{oke90}}
% \tablecomments{}
\end{deluxetable}

\clearpage

\begin{deluxetable}{ll}
\tabletypesize{\tiny}
%\rotate
\tablecaption{Emission and absorption lines\label{emit}}
\tablewidth{0pt}
\tablehead{
 \colhead{Lines} & \colhead{Wavelength ({\AA})}
}
\startdata
Balmer lines & \\
{\hb} & 4862.68 \\
{\ha} & 6564.61 \\
\hline
Forbidden lines & \\
$[${\OII}$]$  & 3727.092, 3729.875 \\
$[${\OIII}$]$ & 4364.436, 4932.60, 4960.295, 5008.240 \\
$[${\OI}$]$   & 6302.046 \\
$[${\NII}$]$  & 6549.86, 6585.27 \\
$[${\SII}$]$  & 6718.29, 6732.67 \\
\hline
Ca H\&K & 3934.777, 3969.588 \\
Mgb     & 5176.7 \\
NaD     & 5895.6 
\enddata
%\tablecomments{Wavelengths are derived from the Algorithm page of
% www.sdss.org/dr7/.}
\end{deluxetable}

\begin{deluxetable}{lllrrrrlll} 
 \tabletypesize{\scriptsize} 
 \tablecaption{Supernova Candidates observed by Subaru/FOCAS \label{sbrspecinfo}} 
 \tablewidth{0pt} 
% \rotate 
 \tablehead{ 
  \colhead{SDSS ID} & \colhead{IAU name\tablenotemark{a}} & \colhead{Type \tablenotemark{b}} &
  \colhead{RA} & \colhead{Dec} & \colhead{$E(B-V)_{MW}$} & 
  \colhead{MJD \tablenotemark{c}} & \colhead{Redshift \tablenotemark{d}} &
  \colhead{Airmass \tablenotemark{e}} & \colhead{$F_{host}/F_{all}$ \tablenotemark{f}}}
 \startdata 
 1119 &  2005fc &  Ia &  -39.58644 &  0.89466 &  0.060 &  53640.26 &  0.2974 (ge) &  1.10 & 0.00,0.09 \\ 
 1166 &  -- &  Ia &  9.35564 &  0.97328 &  0.019 &  53641.39 &  0.3814 (ga) &  1.14 & 0.12,0.00 \\ 
 1686 &  -- &  Ia &  2.24835 &  -0.21039 &  0.055 &  53640.42 &  0.1362 (ge) &  1.07 & 0.41,0.39 \\ 
 1688 &  -- &  Ia &  -38.64223 &  0.32454 &  0.050 &  53641.25 &  0.3591 (ge) &  1.18 & 0.24,0.30 \\ 
 2165 &  2005fr &  Ia &  17.09174 &  -0.09627 &  0.033 &  53641.42 &  0.2841 (sn) &  1.11 & 0.00,0.00 \\ 
 2330 &  2005fp &  Ia &  6.80711 &  1.12060 &  0.024 &  53641.35 &  0.2129 (ge) &  1.27 & 0.34,0.29 \\ 
 2422 &  2005fi &  Ia &  1.99454 &  0.63820 &  0.039 &  53641.32 &  0.2628 (ge) &  1.42 & 0.16,0.20 \\ 
 2661 &  -- &  IIP &  -6.79249 &  0.09721 &  0.038 &  53641.28 &  0.1928 (ge) &  1.50 & 0.52,0.42 \\ 
 2635 &  2005fw &  Ia &  52.70433 &  -1.23808 &  0.096 &  53670.60 &  0.1434 (ge) &  1.39 & 0.45,0.25 \\ 
 2789 &  2005fx &  Ia &  -15.79855 &  0.40110 &  0.051 &  53640.38 &  0.2862 (ga) &  1.06 & 0.00,0.07 \\ 
 2992 &  2005gp &  Ia &  55.49706 &  -0.78267 &  0.087 &  53669.63 &  0.1261 (ge) &  1.56 & 0.79,0.54 \\ 
 3080 &  2005ga &  Ia &  16.93234 &  -1.03950 &  0.048 &  53670.36 &  0.1743 (ge) &  1.10 & 0.36,0.51 \\ 
 3451 &  2005gf &  Ia &  -25.93073 &  0.70815 &  0.056 &  53640.31 &  0.2497 (ge) &  1.10 & 0.19,0.33 \\ 
 3452 &  2005gg &  Ia &  -25.32850 &  0.63919 &  0.059 &  53640.34 &  0.2311 (ge) &  1.06 & 0.35,0.38 \\ 
 5391 &  2005hs &  Ia &  52.34197 &  -1.09470 &  0.111 &  53669.56 &  0.3003 (ge) &  1.18 & 0.04,0.01 \\ 
 5533 &  2005hu &  Ia &  -31.33006 &  0.41343 &  0.070 &  53669.25 &  0.2199 (ge) &  1.06 & 0.21,0.27 \\ 
 5717 &  2005ia &  Ia &  17.89596 &  -0.00589 &  0.031 &  53670.28 &  0.2521 (ge) &  1.44 & 0.25,0.22 \\ 
 5737 &  2005ib &  Ia &  22.85704 &  -0.60339 &  0.038 &  53670.32 &  0.3933 (ge) &  1.25 & 0.39,0.62 \\ 
 5751 &  2005hz &  Ia &  11.63416 &  0.83823 &  0.022 &  53701.33 &  0.1308 (ge) &  1.08 & 0.06,0.65 \\ 
 5844 &  2005ic &  Ia &  -32.21379 &  -0.84294 &  0.121 &  53669.29 &  0.3120 (ge) &  1.09 & 0.16,0.15 \\ 
 5944 &  2005hc &  Ia &  29.19980 &  -0.21372 &  0.030 &  53701.49 &  0.0445 (ge) &  1.84 & 0.12,0.09 \\ 
 5957 &  2005ie &  Ia &  34.76058 &  -0.27283 &  0.036 &  53669.60 &  0.2796 (ge) &  1.86 & 0.00,0.00 \\ 
 6057 &  2005if &  Ia &  52.55362 &  -0.97458 &  0.119 &  53700.54 &  0.0669 (ge) &  1.56 & 0.26,0.29 \\ 
 6108 &  2005ih &  Ia &  1.80660 &  0.34898 &  0.054 &  53671.33 &  0.2599 (ge) &  1.07 & 0.32,0.29 \\ 
 6196 &  2005ig &  Ia &  -22.36885 &  -0.50266 &  0.063 &  53670.25 &  0.2811 (ga) &  1.10 & 0.39,0.41 \\ 
 6249 &  2005ii &  Ia &  3.26555 &  -0.62012 &  0.041 &  53671.39 &  0.2947 (ge) &  1.10 & 0.34,0.40 \\ 
 6406 &  2005ij &  Ia &  46.08861 &  -1.06296 &  0.081 &  53700.50 &  0.1244 (ge) &  1.41 & 0.61,0.63 \\ 
 6471 &  -- &  IIP &  -50.61388 &  0.49495 &  0.075 &  53671.23 &  0.202 (snh) &  1.08 & 0.00,0.00 \\ 
 6699 &  2005ik &  Ia &  -37.18502 &  -1.05699 &  0.053 &  53671.29 &  0.3105 (ge) &  1.12 & 0.04,0.18 \\ 
 9032 &  2005le &  Ia &  -22.11544 &  -0.49357 &  0.060 &  53700.23 &  0.2541 (ge) &  1.17 & 0.00,0.02 \\ 
 9207 &  2005lg &  Ia &  19.08363 &  -0.80780 &  0.038 &  53700.47 &  0.3496 (ge) &  1.82 & 0.46,0.40 \\ 
 10449 &  2005ll &  Ia &  -22.97138 &  -1.12817 &  0.062 &  53701.23 &  0.2415 (ge) &  1.10 & 0.34,0.39 \\ 
 10450 &  -- &  IIn-pec &  -22.72438 &  -1.12867 &  0.064 &  53700.23,53701.27 &  0.5399 (ge) &  1.10 & 0.06,0.02 \\ 
 10550 &  2005lf &  Ia &  -10.32462 &  -1.20487 &  0.036 &  53700.30 &  0.3000 (ge) &  1.19 & 0.08,0.20 \\ 
 12844 &  2006fe &  05gj? &  -46.96182 &  -0.51113 &  0.101 &  53995.25 &  0.0702 (ge) &  1.14 & 0.42,0.36 \\ 
 12855 &  2006fk &  Ia &  -29.74447 &  0.71624 &  0.052 &  53993.27 &  0.1722 (ga) &  1.22 & 0.00,-- \\ 
 12860 &  2006fc &  Ia &  -36.30580 &  1.17591 &  0.069 &  53993.28 &  0.1210 (ge) &  1.19 & 0.13,-- \\ 
 12869 &  2006ge &  Ia &  -32.60088 &  0.00108 &  0.140 &  53996.24 &  0.2841 (ge) &  1.15 & 0.07,0.12 \\ 
 12883 &  2006fr &  Ia &  -47.37411 &  0.39867 &  0.102 &  53995.30 &  0.3060 (ge) &  1.06 & 0.53,0.32 \\ 
 12972 &  2006ft &  Ia &  7.95859 &  -0.38296 &  0.023 &  53995.43 &  0.2606 (ge) &  1.09 & 0.00,0.00 \\ 
 12977 &  2006gh &  Ia &  13.69561 &  -0.25086 &  0.025 &  53996.50 &  0.2474 (ge) &  1.08 & 0.12,0.19 \\ 
 12979 &  2006gf &  91bg? &  11.60148 &  0.00348 &  0.020 &  53996.46 &  0.1164 (ga) &  1.06 & 0.41,0.22 \\ 
 12991 &  2006gd &  IIP &  17.60995 &  -1.06803 &  0.073 &  53996.53 &  0.1539 (ge) &  1.12 & 0.19,0.08 \\ 
 13000 &  -- &  IIn &  20.74847 &  -0.63447 &  0.037 &  53995.58 &  0.3513 (ge) &  1.25 & 0.34,0.07 \\ 
 13025 &  2006fx &  Ia &  -18.43267 &  0.41590 &  0.083 &  53995.35 &  0.2240 (ge) &  1.09 & 0.29,0.16 \\ 
 13099 &  2006gb &  Ia &  -0.18127 &  -1.25038 &  0.039 &  53995.39 &  0.2655 (ge) &  1.12 & 0.32,0.30 \\ 
 13152 &  2006gg &  Ia &  7.05211 &  0.11801 &  0.021 &  53996.40 &  0.2031 (ge) &  1.13 & 0.35,0.15 \\ 
 13174 &  2006ga &  Ia &  13.23469 &  0.44786 &  0.026 &  53995.48 &  0.2357 (ge) &  1.06 & 0.30,0.23 \\ 
 13327 &  2006jf &  Ia &  -21.27261 &  0.00204 &  0.057 &  53996.32 &  0.2814 (ge) &  1.13 & 0.55,0.47 \\ 
 13346 &  -- &  AGN &  -20.86334 &  0.32819 &  0.066 &  53996.36 &  1.3716 (agn) &  1.06 & --,-- \\ 
 13376 &  2006gq &  II? &  26.30225 &  0.35644 &  0.037 &  53996.60 &  0.0696 (ge) &  1.29 & 0.15,0.26 \\ 
 13934 &  2006jg &  Ia &  -17.88899 &  -0.43527 &  0.102 &  54023.41 &  0.3310 (ge) &  1.33 & 0.12,0.41 \\ 
 14261 &  2006jk &  Ia &  -31.75960 &  0.25379 &  0.094 &  54023.27 &  0.2853 (ge) &  1.06 & 0.00,0.00 \\ 
 14298 &  2006jj &  Ia &  -45.10495 &  1.22326 &  0.076 &  54023.23 &  0.2692 (sn) &  1.06 & 0.00,0.00 \\ 
 14456 &  2006jm &  Ia &  -16.44890 &  1.05066 &  0.090 &  54023.46 &  0.3283 (ge) &  1.77 & 0.28,0.35 \\ 
 14475 &  -- &  hyp? &  -23.87125 &  0.20341 &  0.084 &  54023.30 &  0.1437 (ge) &  1.06 & 0.27,0.21 \\ 
 15170 &  2006jx &  91T? &  58.05858 &  0.29207 &  0.245 &  54023.57 &  0.3999 (sn) &  1.11 & --,0.06 \\ 
 15217 &  2006jv &  Ia &  22.63421 &  0.21980 &  0.029 &  54023.49 &  0.371 (sn) &  1.18 & 0.00,0.00 \\ 
 15219 &  2006ka &  Ia &  34.61115 &  0.22669 &  0.037 &  54023.54 &  0.2467 (ge) &  1.23 & 0.13,0.14 \\ 
 16631 &  2006pv &  Ia &  54.62279 &  -0.67557 &  0.128 &  54065.53 &  0.207 (sn) &  1.55 & --,0.37 \\ 
 16758 &  2006pw &  Ia &  -18.38765 &  1.15081 &  0.085 &  54065.25 &  0.3261 (ge) &  1.12 & 0.02,0.01 \\ 
 16776 &  2006qd &  Ia &  -12.97859 &  -0.17608 &  0.044 &  54066.30 &  0.2667 (ge) &  1.31 & 0.32,0.37 \\ 
 16779 &  2006qa &  Ia &  45.29054 &  -0.02624 &  0.105 &  54065.44 &  0.3983 (ge) &  1.16 & 0.59,0.04 \\ 
 16781 &  2006qb &  Ia &  52.06103 &  -0.20096 &  0.111 &  54065.48 &  0.3257 (ge) &  1.29 & 0.01,0.01 \\ 
 16840 &  2006qf &  II? &  57.98849 &  -0.39658 &  0.160 &  54065.56 &  0.1662 (ge) &  1.85 & --,0.07 \\ 
 16847 &  2006px &  Ia &  -6.67875 &  0.96110 &  0.033 &  54066.30 &  0.2767 (ge) &  1.44 & 0.24,0.28 \\ 
 16938 &  2006qe &  Ia &  -11.91674 &  -0.53442 &  0.047 &  54065.29 &  0.386 (sn) &  1.19 & 0.61,0.12 \\ 
 16953 &  2006pp &  Ia &  12.24364 &  0.47695 &  0.024 &  54065.33 &  0.3387 (ge) &  1.11 & 0.27,0.19 \\ 
 17048 &  2006qi &  Ia &  27.58151 &  0.88490 &  0.025 &  54065.36 &  0.189 (sn) &  1.10 & 0.19,0.38 \\ 
 17081 &  2006ql &  Ia &  27.53797 &  0.42048 &  0.038 &  54065.40 &  0.2749 (ga) &  1.21 & 0.02,0.42 \\ 
 17117 &  2006qm &  Ia &  40.60031 &  -0.79652 &  0.032 &  54066.38 &  0.1404 (ge) &  1.08 & 0.00,0.00 \\ 
 \enddata 
\tablecomments{RA and Dec in Columns 4 and 5 are the Right Ascension and Declination (J2000) in degrees.}
\tablenotetext{a}{IAU names were not attached to five SNe due to on-site analysis.}
\tablenotetext{b}{``05gj?", ``91bg?", ``hyp?", ``91T?" indicates that the object resembles 2005gj-like SN Ia, 1991bg-like SN Ia, hypernova, 1991T-like SN Ia. }
\tablenotetext{c}{The observational mid-date. The red part of a spectrum was first observed and the blue part followed for most of the cases.}
\tablenotetext{d}{Heliocentric redshift. The redshift of the AGN (SN13346) was measured from CIII] 1908.73 {\AA}, {\MgII} 2797.92 {\AA} and two {\OII} 3726.03,3728.82 {\AA}. ``ge'' indicates that the redshift of the target was determined from galaxy emission line(s). ``ga'' from galaxy absorption line(s), ``sn'' for redshift from the SN spectrum fitting and ``snh'' from the {\ha} emission line of the SN IIP.}
\tablenotetext{e}{Mid airmass. The red part of a spectrum was first observed and the blue part followed for most of the cases.}
\tablenotetext{f}{Host contaminations for blue and red side spectrum, respectively.}
\end{deluxetable}

\begin{deluxetable}{lll}
 \tabletypesize{\tiny}
 \tablecaption{Observational Summary of FOCAS Followup Campaign \label{obsf}} 
 \tablewidth{0pt}
 \tablehead{
 \colhead{Classification} & \colhead{Number} & \colhead{Comments}
 }
 \startdata
 Supernova & & \\ 
 - normal Ia    & 59 ($=31+28$) & 2005/2006yr incl. non-IAU named Ia (3$^a$) \\
 - peculiar Ia  & 3 & SN12844$^b$, SN12979$^c$, SN15170$^d$ \\
 - type IIn     & 7 & SN10450 (peculiar IIn) \\
 Probable Hypernova & 1 & SN14475$^e$ \\
 AGN            & 1 & SN13346 \\
 \enddata
 \tablenotetext{a}{SN1166, SN1686, SN1688}
 \tablenotetext{b}{SN 2005gj-like Spectrum. Large {\csalt} value.}
 \tablenotetext{c}{SN 1991bg-like Spectrum. The $ri$-band lightcurves are
 well-fitted by the SALT2 code, however, the $g$-band brightness seems almost
 constant during 40 days after the $B$-band maximum date.}
 \tablenotetext{d}{SN 1991T-like spectrum. The lightcurve is normal.}
 \tablenotetext{e}{Peculiar spectrum. Large {\csalt} value.} 
\end{deluxetable}

\clearpage

\end{document}